Fully denaturing two-dimensional electrophoresis of membrane proteins: a critical update.

Thierry Rabilloud 1,2, Mireille Chevallet 1,2, Sylvie Luche 1,2, Cécile Lelong 1,2

1 CEA-DSV/iRTSV/LBBSI, Biophysique et Biochimie des Systèmes Intégrés, CEA-Grenoble, 17 rue des martyrs, F-38054 GRENOBLE CEDEX 9, France

2 CNRS UMR 5092, Biophysique et Biochimie des Systèmes Intégrés, CEA-Grenoble, 17 rue des martyrs, F-38054 GRENOBLE CEDEX 9, France

Correspondence :
Thierry Rabilloud, iRTSV/BBSI
CEA-Grenoble, 17 rue des martyrs,
F-38054 GRENOBLE CEDEX 9
Tel (33)-4-38-78-32-12
Fax (33)-4-38-78-44-99
e-mail: Thierry.Rabilloud@ cea.fr

Abstract

The quality and ease of proteomics analysis depends on the performance of the analytical tools used, and thus of the performances of the protein separation tools used to deconvolute complex protein samples. Among protein samples, membrane proteins are one of the most difficult sample classes, because of their hydrophobicity and embedment in the lipid bilayers. This review deals with the recent progresses and advances made in the separation of membrane proteins by two-dimensional electrophoresis separating only denatured proteins. Traditional 2D methods, i.e.methods using isoelectric focusing in the first dimension are compared to methods using only zone electrophoresis in both dimensions, i.e. electrophoresis in the presence of cationic or anionic detergents. The overall performances and fields of application of both types of method is critically examined, as are future prospects for this field

## 1. A historical introduction: the protoproteomics era

Because of their strategic localization at the interface between the cell and its external environment, which impart for many of their roles (transport, sensing, communication), membrane proteins have received a lot of attention from the entire field of biochemistry, and proteomics makes no exception to this rule.

As a matter of facts, the first attempts of separation of membrane proteins by 2D gels quickly followed the first detailed descriptions of 2D electrophoresis [1]. Several modifications of the basic 2D electrophoresis protocol were published in the 80's and early 90's, each being described as "optimized" for membrane proteins, but following the basic constraints in protein solubilization [2]. As expected from membrane protein chemistry, these protocols varied mainly by the detergent used in the IEF dimension, as this is a crucial point for the total 2D electrophoresis protocol. Compared to the initial 2D electrophoresis protocols, which used NP-40 or Triton X100 in combination with urea as the protein solubilization agent, these "improved" protocols used a variety of non-ionic or zwitterionic detergents, including CHAPS [3], linear sulfobetaines [4], amidosulfobetaines [5] or dodecylmaltoside [6]. However, most of these papers were poorly demonstrative, as they just relied on presence of additional spots in the improved system to claim for solubilization of membrane proteins. It must be kept in mind that protein identification means at that time were very far from what they are now, especially for direct protein identification. The most sensitive protein identification means was protein immunoblotting, but this realizes a targeted identification (where is protein X) rather than a naive identification (what protein is in this spot).

Despite this important difficulty, some direct evidence of the solubilization of membrane proteins could already be gained during this period. Demonstration of analysis by 2D electrophoresis of transferrin receptor [7] or ACTH receptor [8] in complex membrane samples was obtained by immunoblotting. Conversely, analysis by 2D electrophoresis of semi-purified membrane preparations allowed to identify some membrane receptors by classical staining [9-11]. But in a few cases, definitive evidence of poor performance of classical 2D electrophoresis protocols with well-known membrane proteins could also be established [12], and the overall situation of the performance of 2D electrophoresis for membrane proteins using these classical protocols based on urea-detergent as the solubilization agent in IEF has been reviewed [13].

2. The proteomics revolution: taking the measure of the problem

The advent of ultrasensitive protein identification, first by Edman sequencing then by mass spectrometry (with a further important increase in sensitivity), have had a major impact on the whole protein biochemistry field. In the field of the wide-scale analysis of membrane proteins, this revolution was even more dramatic in its impact, as the community became aware rather fast of an important problem in the analysis of membrane proteins by 2D electrophoresis. Within a few years, it became obvious that most of the spots visualized on 2D gels of membrane preparations were mostly soluble contaminants, extrinsic proteins, and that only very few intrinsic membrane proteins, defined as proteins with one or several transmembrane helices, were present on those 2D gels [14, 15]. Careful examination of the physico-chemical features of the proteins identified on 2D gels revealed that the general hydropathy of the polypeptide chain had a major impact on its ability to be seen in 2D gels [16]. As intrinsic membrane proteins are generally more hydrophobic than classical intracellular proteins, this explained at least in part why intrinsic membrane proteins were so poorly represented on classical 2D gels. However, more detailed examination of the features of membrane proteins separated on 2D gels revealed that the average hydropathy index (GRAVY) is not the best predictive index for visualization on 2D gels, and that the ratio between the number of predicted transmembrane domains over molecular weight [13] is a better predictive index.

It seems rather obvious that the main problem encountered is the solubilization of the hydrophobic membrane proteins under the conditions prevailing in the IEF dimension (low ionic strength, no ionic detergent). However, several experiments suggest that besides this solubilization problem, there is a protein detection problem for membrane proteins. This was clearly suggested by a publication on brain membrane proteins [17], but this can also be seen on figure 1.

However, in this latter example of inner mitochondrial membrane transporters, there is clearly a superposition of a detection problem and of a solubilization problem. The detection problem is highlighted by the fact that some membrane proteins are detection by the MS-incompatible protocol and not by the MS-compatible protocol. In this case, excision on the gels stained with the MS-compatible protocol of gel pieces that are unstained but superposable to stained areas in the gel stained with the MS-incompatible protocol led to positive protein identification (e.g. ADT2, SFX3). This is indicative of a detection problem. However, excision of other unstained gels pieces, in the areas where other missing transporters (e.g. phosphate transporter, glutamate-malate shuttle) should be (from their calculated pI and Mw), did not yield any additional identifications, showing that protein solubility in 2D gels is still a key issue, and needed to be improved.

3. Improvements of the standard 2D electrophoresis technique

The constraints induced by the IEF separation are rather strict: low ionic strength, no ionic detergent in the gel, and low amounts of ionic detergent in the sample. Consequently, there are only two parameters on which the experimentator can play with to increase the solubility o proteins are the nature and concentrations of chaotropes and the nature and concentrations of detergents. Both the chaotrope and detergent can be a single compound or a mixture of various compounds. The historical chaotrope in IEF is urea, as this is an efficient one and the only one to be compatible with acrylamide polymerization. However, the above-mentioned corpus of knowledge had shown that urea alone, whatever nonionic detergent it was used with, was poorly efficient for the solubilization of membrane proteins.

The situation was improved with the introduction of thiourea as an ancillary chaotrope in addition to urea [19]. While this initial report was purely qualitative and not related to membrane proteins, dedicated studies investigating several types of detergents in combination with a urea-thiourea chaotrope were carried out on various membrane systems. As a matter of facts, the change in chaotrope dramatically affected the solubilizing power of even "mild" non ionic detergents, as shown by the work carried out on fat globules in human milk [20]. While the combination of urea and Triton X100 leaded to poor patterns, a combination of urea-thiourea and Triton X 100 solubilized and focused human butyrophilin, which is a protein with a transmembrane domain.

This chaotropic combination was tested with various types of detergents, including amidosulfobetaines [21] and other types of sulfobetaines [22], [23]. For this class of detergents, a detailed structure-efficiency study was carried out [24], and solubilization of bona fide intrinsic membrane proteins was demonstrated, including GPCR with seven transmembrane domains [17], red blood cell membrane proteins with various transmembrane domains, including the Band III protein with 12 transmembrane domains [22] which was reluctant to solubilization so far [12], or aquaporins and proton ATPase [21], [23], or mitochondrial transporters [25].

However, sulfobetaines were not the only type of detergent which proved useful. Nonionic detergents of the oligooxyethylene group proved efficient on the same proteins [26]. Detergents of the glucoside type also proved useful [26], and were able to solubilize transmembrane proteins from myelin [27]. Finally zwitterionic detergents of the phosphocholine type were also tested and able to solubilize membrane proteins from muscle [28].

These positive results prompted several groups to investigate the possibility to analyze membrane proteins by 2D electrophoresis on more complex systems, either from plants (e.g. [29]) or on animals (e.g. [30, 31]). However, it is fair to say that the results were generally considered as disappointing. While it is true that the improved methods did allow to visualize some membrane proteins (e.g. in [32]), the general situation is that many membrane proteins are missing [33]. While this impression was first based on previous knowledge (e.g. in [33]), the concomitant use of proteomics strategies not based on 2D gels, such as shotgun strategies [34] or strategies based on 1D SDS gels [35], made obvious that many hydrophobic proteins are missing on classical 2D gels [36]. To sum up the situation with the example of the P450 cytochromes, the improvements made on protein solubilization for IEF allowed to go from none [37] to some (see figure 2), but these are obviously too few.

A happy exception to this rule lies in the bacterial membrane proteins, and especially in the proteins of the outer membrane (OMPs) of Gram- bacteria [40]. These proteins are essentially of the porin type, and their transmembrane part is not made of helices but of a beta barrel. Once properly denatured, these proteins are fairly soluble in the conditions prevailing in IEF, and can thus be analyzed with high resolution. this is true for bacterial porins [41] but also for eukaryotic porins [42].
Consequently, proteomics based on 2D gels has been widely applied to bacterial membrane proteomics. While the success has been limited for Gram+ bacteria [43, 44], where there is only one membrane with most proteins spanning the membrane by helices, much more work has been devoted to Gram- bacteria, with greater success such as the identification of a new OMP [45], the study of E. coli either from the basic microbiology point of view [46] or from the side pathogenic bacterial strains [47, 48], and more generally the study of various Gram- organisms of interest in various areas [49-57].

However, this exception shall not mask the general rule, which is that proteins having multiple transmembrane helices generally escape analysis by IEF-based two-dimensional electrophoresis [58]. The presence of lipids was claimed to be deleterious [59], but delipidation with organic solvents did not induce a major improvement in the solubilization of membrane proteins in classical 2D PAGE [33]. It was soon demonstrated that isoelectric precipitation is the major phenomenon to blame [60], which led to the attempt to solubilize the isoelectrically-precipitated proteins with hot SDS [61]. Unfortunately, this method did not really solve the problem, so that other solutions had to be sought.

4. 2D electrophoresis without IEF

As it appears that limited solubility of the membrane proteins is the main limiting factor for their analysis by classical 2D PAGE, alternate methods must be found. In this respect, the contrast between the poor solubilizing performances under IEF conditions and the excellent ones in SDS PAGE gives valuable insights into the directions to be followed for dedicated solubilization systems for electrophoretic separation of membrane proteins. Of course, membrane proteins must be separated in the presence of detergents to cover the hydrophobic parts of the protein. But in addition, the electrostatic repulsion between molecules must be maximal to prevent aggregation. To this purpose, it is often advantageous to add extraneous charges to proteins via a charged protein-binding agent. Consequently, two main types of 2D gels can be used for separating membrane proteins:

In the first type, the first dimension uses native electrophoresis of membrane proteins and/or membrane complexes, generally with a charge-modifying agent. This concept, which traces back very early in electrophoresis and has been reviewed previously [62], has been refined and further developed more recently [63] and will be reviewed in another article of this issue [64].

In the second type, the first dimension separates denatured proteins, and in this case, the denaturing agent is most often also the charge transfer agent, and is made of an ionic detergent. Of course, it would be of little interest to use twice the simple SDS-PAGE technique, as the proteins would simply lay on the diagonal of the gel. Thus, the optimal system in the first dimension should offer a separation as different as possible from SDS PAGE, while keeping the high loading capacity and high solubilizing power of SDS PAGE.

Among the various electrophoretic systems designed to date, the urea-16BAC system originally devised by MacFarlane [65] has these desirable features. It shows a high loading capacity [66], while also showing a very different migration when compared to SDS [67]. Its ability to separate bona fide integral membrane proteins and thus its utility in membrane proteomics was demonstrated rather early [68], and when it became obvious that IEF-based 2D gels did not show adequate solubilization performances, this system became an obvious choice, as shown for example on bacterial membrane proteins [69]. It therefore received a lot of applications in various fields, spanning from bacterial proteins [70, 71], to yeast [72], to mammalian cells or tissues [73, 74] or to subcellular membranes [75].

Thus, this system, as well as closely related ones using other cationic detergents such as CTAB instead of 16-BAC [76] have gained increased popularity in membrane proteomics. Compared to the initial description [65], the newest versions do not use the staining between the two dimensions, but in a simple equilibration in the SDS-containing buffer [76]. However, the cationic system is not as straightforward to use and not as versatile as SDS PAGE.

For example, the polymerization of gels at low pH cannot be achieved by the classical and robust TEMED/persulfate system, and the more delicate ascorbate/ferrous ion/hydrogen peroxide system is often used. Alternatively, the methylene blue-based photopolymerization system [77], which has been shown to be efficient with other types of acidic, urea-containing gels [78], can be used to polymerize the acidic first-dimension gels [79].

In addition, and oppositely to SDS PAGE, the performance of the urea-cationic detergents systems is very sensitive to the pH of the separating gels, as shown on figure 3.

This suggests that the solubilization is not fully driven by the cationic detergent, but rather both by the native charge of the proteins and the charges added by the detergent. This suggests in turn that the cationic detergents are less efficient than SDS for the solubilization of proteins, so that it can be expected that some membrane proteins are soluble in SDS-containing media and not in cationic detergents-containing media. This assumption has recently been demonstrated as true [82].

Thus, if optimal membrane protein solubilization is required, SDS must be used in both dimensions. Consequently, some tricks must be found to reach somewhat different separations in the two SDS-based separations. This can be achieved to some extent by changing the buffer system from one dimension to another, as this alters the resolution of the system [83]. To further enhance this differential mobility effect, nonionic modifyers can be used, such as glycerol [84] or urea [85]. Both approaches have been shown to solubilize adequately membrane proteins [82], [86], and the urea-based approach has been shown to be superior to the cationic detergents-based approaches in terms of hydrophobic proteins solubilization [82].

However, this comparison also showed that the resolution on 2D gels, in terms of spot spreading on the gels, clearly ranges in the order: double SDS techniques < cationic/anionic detergent techniques << IEF-based techniques. While this was easily predictable on the basis of the interdependency of the separation principles used for the two-dimensions of the system, this is not without consequences on how we can use these electrophoretic systems optimally in proteomics studies and on how we can improve their use.

5. Future prospects

One of the great strengths of classical, IEF-based 2D electrophoresis is its very high resolution. It allows to separate many post-translational variants, to separate spots enough so that image analysis is feasible to follow spots variations between various conditions investigated, and finally to make the basic assumption that one spot contains most often a single protein, or at least a single dominating protein, so that the spot volume changes can be attributed to the change in abundance of this protein.

When switching to 2D systems based only on detergent zone electrophoresis, the loss of resolution impacts the various features quite differently. While the ability of separating simple post-translational variants is irremediably lost because of the separation principles at play, the factors linked to spot crowding are impacted both by the resolution of the gel system and by the complexity of the sample.

Positional variability, i.e. the difficulties encountered because of gel to gel variations in spot positions, are of course more severe on crowded gels where only a fraction of the gel space is used to display proteins. These difficulties can be dealt with by multiplexing, i.e. by labeling several samples with different fluorescent probes and by comigrating them in a single gel [87]. this approach has been applied successfully to cationic/anionic 2D PAGE (e.g. in [88] ).

However, the other problem linked to spot crowding, i.e. fusion of several proteins in a single, average spot, cannot be solved easily, so that we are currently facing a difficult situation. So-called membrane preparations contain both intrinsic membrane proteins and more soluble proteins associated to membrane by various mechanisms. As a matter of facts, intrinsic membrane proteins often represent a low percentage (at least in mass) of the proteins present in the biochemical preparation, and conversely most of the proteins present in the preparations are soluble ones.

Consequently, the dilemma that we are facing now is quite simple: we can analyze easily the soluble components of a membrane preparation by classical 2D electrophoresis, but most of the scarce resolution space that we have in double-zone 2D PAGE is wasted to analyze those soluble components, and little is left to analyze the intrinsic proteins.

Thus, our efforts to increase the total performance of proteomics of intrinsic membrane proteins should go into two directions, i) finding double-zone 2D PAGE systems with increased resolution and ii) finding biochemical ways to enrich preparations in intrinsic membrane proteins. However, both directions are likely to be difficult to improve.

As to the electrophoresis systems, we are bound by the protein-binding capacities of detergents. A detergent showing a very different protein binding compared to SDS is desirable to induce a different migration, However, when this is the case, it is also likely that the detergent/protein ratio will be rather low, resulting in lower solubilizing performances, especially at the high protein concentrations prevailing in electrophoresis. Conversely, a good protein solubilizer will bind at a high detergent/protein ratio, thereby leading to a migration resembling the one shown in SDS and therefore to limited off-diagonal effects.

As to the enrichment in intrinsic membrane proteins, several solutions have been investigated, with limited success and robustness. The most widely used enrichment consists of washing membranes in high salt and/or high pH solutions. However, this leads to very limited enrichment in intrinsic membrane proteins, which represent ca. 10 percent of the proteins in the unwashed membranes and ca. 20 percent in the washed ones [89]. Enrichment by two-phase partitioning has also been described, but the real gain in performance has not been quantified [90].

Other strategies are based on the selective extraction or purification of defined classes of proteins. For example, classes of surface glycoproteins have been purified prior to proteomic analysis [91]. Besides the fact that the unglycosylated membrane proteins are lost by such an approach, the performances in terms of efficiency, i.e. number of missed glycoproteins / number of selected glycoproteins, are difficult to evaluate.

A direct and straightforward approach could be to select the intrinsic membrane proteins on the basis of their hydrophobicity, e.g. by extracting them into organic solvents [92, 93]. Although this method clearly achieves a major enrichment in hydrophobic proteins [93], some more soluble, non membrane proteins are also extracted [92], and the methods has an opposite bias when compared to alkaline washes: while alkaline washes let a lot of non membrane proteins go with the membrane fraction, organic extraction leaves a lot of membrane proteins in the pellet with soluble ones.

Sometimes, a combination of electrophoretic approaches can be used to further increase the representation of membrane proteins. For example, membrane supercomplexes can be first isolated by native electrophoresis [64] and the isolated supercomplexes can be then analyzed by high-resolution elelctrophoresis, as shown in [94]

6. Concluding remarks

Because of the high interest in membrane proteins, the proteomics of membrane proteins is still a hot topic in proteomics, as shown in a recent review in the field [95]

Compared to the initial hopes, it is fair to say that two-dimensional electrophoretic separation of membrane proteins did not show adequate performances, as there seems to be an inverse correlation between resolution and solubilizing power. Consequently, current 2D protocols do not handle adequately the complexity of whole membrane preparations. This does not mean that this type of separation is useless, and the numerous references cited here clearly show that adapted 2D separations, or even better a combination of approaches such as in [96] can indeed lead to the identification of some membrane proteins. However, it is quite clear that we do not have for the moment the same degree of performance, general applicability and robustness on membrane proteins that we have on cytosolic proteins with classical 2D PAGE, and that further research both in electrophoretic system and in membrane protein enrichment is needed to improve the situation to a level of general application.

Figures

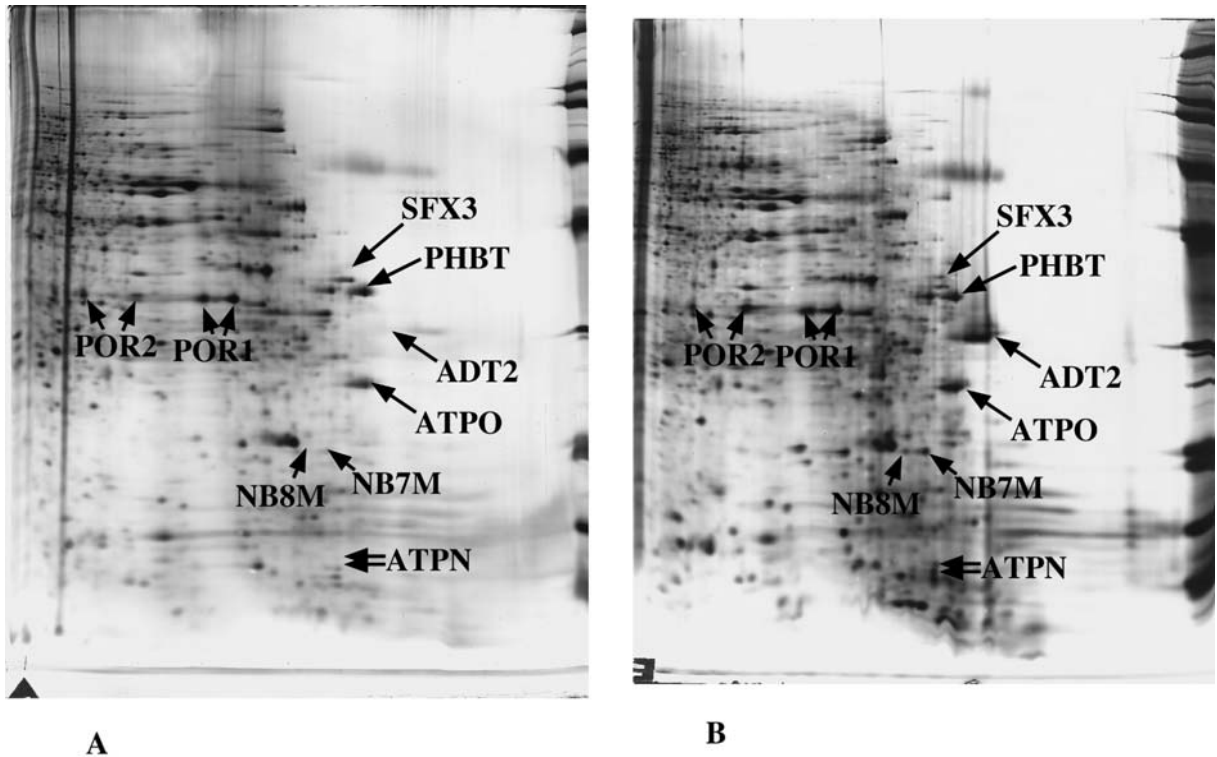

Figure 1: Staining artefacts in membrane proteins

Total mitochondrial proteins from human placenta (150 µg) were separated by two dimensional electrophoresis. The IPG gradient is a homemade 5.5-12 linear pH gradient, and the proteins have been extracted and focused in a 7M urea, 2M thiourea, 2% Brij56 and 0.4% pharmalytes 3-10 mixture. sample loading by anodic cup loading. The proteins were stained with silver, using either a mass spectrometry compatible silver ammonia protocol [18] (panel A), or the same protocol where the initial acid alcohol fixation is replaced by a 4% formaldehyde 20% ethanol fixation for 1 hour (panel B).

Protein identifications were carried out by MALDI peptide mass fingerprinting on the (A) gel.

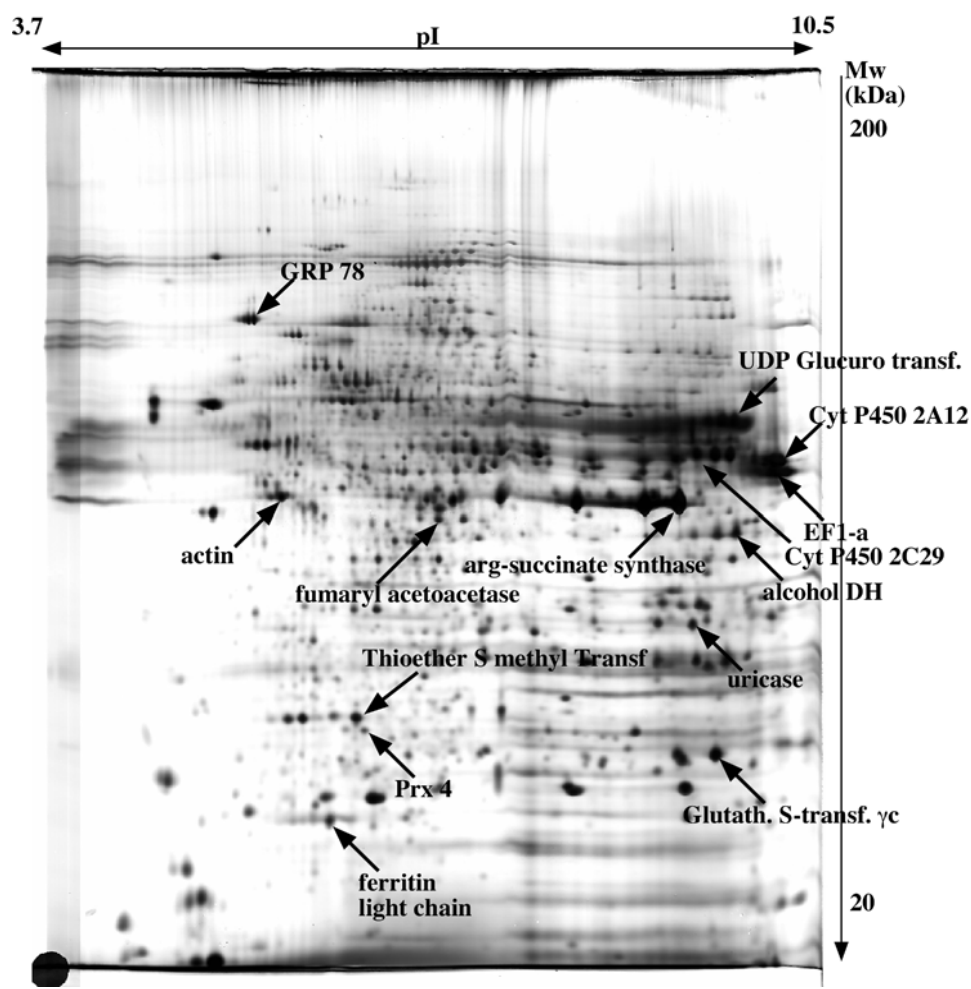

Figure 2: Two dimensional electrophoresis of mouse microsomal proteins
Proteins from mouse liver microsomes were separated by two-dimensional electrophoresis. The IPG gradient is a homemade 3-10.5 linear pH gradient [38], and the proteins have been extracted and focused in a 7M urea, 2M thiourea, 2% Brij56 and 0.4% pharmalytes 3-10 mixture. sample loading (150µg) by in gel rehydration. Silver staining by silver nitrate staining [39] with formaldehyde developer.
Protein identifications were carried out by MALDI peptide mass fingerprinting.
Note the presence of cytochrome P450 isoforms in the 2D gel.

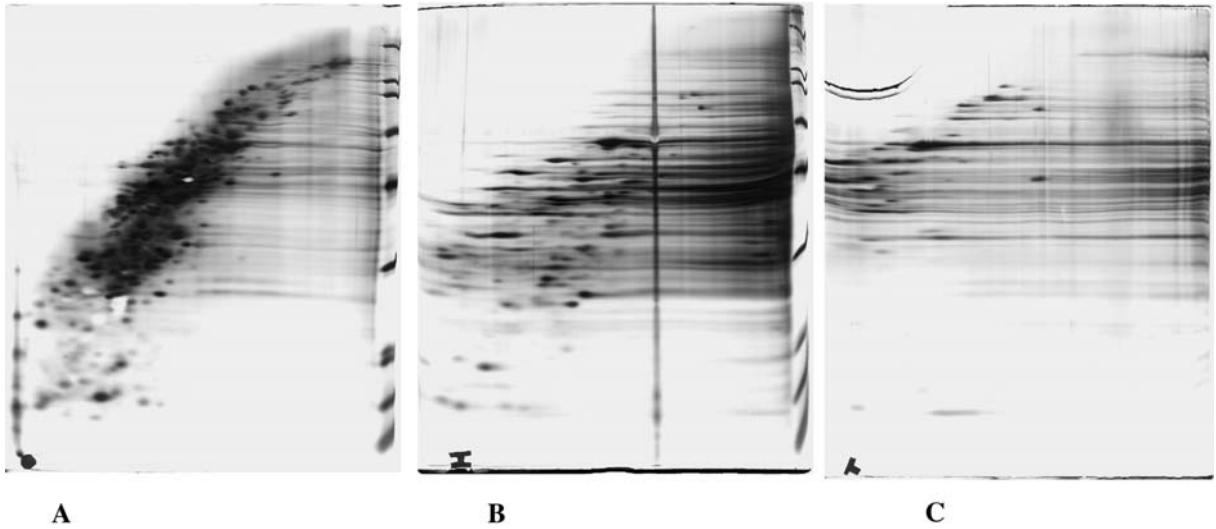

Figure 3: cationic/anionic 2D PAGE of bacterial membrane proteins

Membrane proteins from B. subtilis (50µg) were separated by 16BAC/SDS PAGE. The first dimension, 16BAC gels were run either at pH 2 in the phosphate/glycine system [65] (panel A), or at pH 5 in the acetate/beta alanine system [80] (panel B), or at pH 7 in the Hepes/histidine system [81] (panel C). protein detection by silver ammonia staining [18]

Note the increased streaking with increasing running pH